\documentclass [12pt]{article}
\usepackage{graphicx,epsfig}
\title{Ground-state properties of fermionic mixtures with mass imbalance 
in optical lattices}
\author{Pavol Farka\v sovsk\'y \\
Institute of Experimental Physics, Slovak Academy of Sciences\\
Watsonov\'a 47, 040 01 Ko\v sice, Slovakia}
\date{}
\begin{document}
\baselineskip=20pt
\maketitle

\begin{abstract}
Ground-state properties of fermionic mixtures confined in a
one-dimensional optical lattice are studied numerically within 
the spinless Falicov-Kimball model with a harmonic trap. 
A number of remarkable results are found. (i) At low particle 
filling the system exhibits the phase separation with
heavy atoms in the center of the trap and light atoms in the surrounding 
regions. (ii) Mott-insulating phases always coexist with metallic phases.
(iii) Atomic-density waves are observed in the insulating regions for 
all particle fillings near half-filled lattice case.
(iv) The variance of the local density exhibits the universal behavior
(independent of the particle filling, the Coulomb interaction and the
strength of a confining potential) over the whole region of the local density 
values.  
\end{abstract}

\newpage
\section{Introduction}
The ability to confine ultracold Bose and Fermi gases inside artificial crystals 
generated by standing-wave laser light fields, i.e., optical lattices, offers 
the possibility to create ideally clean and highly tunable strongly interacting 
quantum many-body systems~\cite{Greiner}. The low-energy properties of these 
systems can be described using models borrowed from condensed matter
systems~\cite{Cirac} whose parameters and dimensionality can be controlled 
with high precision. This fact opens new routes for understanding the physics 
of long-standing problems of strongly correlated systems, like the phase 
separation, metal-insulator transitions, superconductivity, etc. On the other 
hand, the quadratic confining potential, present in addition to the regular 
"lattice" potential leads to a number of fundamentally new phenomena. 
For example, it was shown~\cite{Rigol1,Rigol2} that in the presence 
of a confining potential the Mott-insulating phase is restricted to the domain
that coexists with a metallic phase, in contrast to the global character 
typical of solid-state systems. Moreover, mixtures of two-component atoms with
different masses (e.g., $^6Li$ and $^{87}Rb$) introduce an additional parameter, 
namely, the difference between the hopping amplitudes associated with each 
species of atoms in the optical lattice. This may affect the stability of 
the possible quantum phases or even induce new ones~\cite{Ates,Gu,Caza}. 
For these reasons the mixtures of ultracold atoms in optical lattices belong 
to the most intensively studied subjects of contemporary experimental 
and theoretical physics. 

In this paper we investigate the ground-state properties of mixtures of two 
species of fermionic atoms with strongly different masses in a harmonic 
potential. Since the tunneling rate decreases exponentially with the square
root of mass of the atom, and particles of different species on the optical 
lattice interact only through the on-site interaction, this leads naturally 
to the Falicov-Kimball Hamiltonian~\cite{Falicov} with a confining
potential~\cite{Rigol1} 
\begin{equation}
H=\sum_{ij}t_{ij}d^+_id_j + U\sum_i d^+_id_if^+_if_i 
+ \left(\frac{2}{L} \right)^2V\sum_i\left(i-\frac{L}{2}\right)^2(d^+_id_i+f^+_if_i) 
\label{eq1}
\end{equation}
where $f^+_i$($f_i$) and $d^+_i$($d_i$) are the creation (annihilation) 
operators of heavy ($f$) and light ($d$) particles at lattice site $i$. 
The number of lattice sites is $L$ and is selected so that all the 
fermions are confined in the trap. We denote the total number of fermions 
in the trap as $N$ and consider equal number of heavy ($N_f=\sum_if^+_if_i$)
and light ($N_d=\sum_id^+_id_i$) atoms.

The first term of Eq.~1 is the kinetic energy corresponding to 
quantum-mechanical hopping of the light $d$ atoms between the nearest-neighbor
sites $i$ and $j$. These intersite hopping transitions are described by the 
matrix  elements $t_{ij}$, which are $-t_d$ if $i$ and $j$ are the nearest 
neighbors and zero otherwise (in the following all parameters are measured
in units of $t_d$). The second term represents the on-site Coulomb interaction 
($U>0$) between the light and heavy atoms. 
The last term is the energy of light and heavy atoms in the harmonic trapping 
potential. In accordance with similar studies for the asymmetric Hubbard
model~\cite{Gottwald}, we consider here the same trapping potential for 
both species of atoms.

Since in this spinless version of the Falicov-Kimball model
with a confining potential the $f$-heavy atom occupation
number $f^+_if_i$ of each site $i$ commutes with
the Hamiltonian (1), the $f$-heavy atom occupation number
is a good quantum number, taking only two values: $w_i=1$
or 0, according to whether or not the site $i$ is occupied
by the heavy atom. Therefore, the Hamiltonian (1) 
can be written as

\begin{equation}
H=\sum_{ij}h_{ij}d^+_id_j+
\left(\frac{2}{L}\right)^2V\sum_i\left(i-\frac{L}{2}\right)^2w_i,
\end{equation}
where $h_{ij}(w)=t_{ij}+
\left(Uw_i +\left(\frac{2}{L}\right)^2V\left(i-\frac{L}{2}\right)^2\right)
\delta_{ij}$.

Thus for a given configuration of heavy atoms
$w=\{w_1,w_2 \dots w_L\}$ defined on the one-di\-men\-sional
lattice, the Hamiltonian (2) is the second-quantized version of 
the single-particle Hamiltonian $h(w)$, so the investigation of
the model (2) is reduced to the investigation of the
spectrum of $h$ for different configurations of heavy atoms.

It is well known that in the absence of harmonic confinement ($V=0$) the 
one-dimensional Falicov-Kimball model exhibits a rich spectrum of solutions 
including phase separated ($U$ small) as well as most homogeneous 
distributions of heavy atoms~\cite{Gruber}.
However, due to the confining potential the lattice sites become, inequivalent,
and thus, it is of fundamental importance to analyse the interplay between 
the on-site Coulomb interaction and the confining potential. 

To describe the system at nonzero 
$V$ we have calculated various local quantities, like 
the local density of heavy atoms ($n^f_i=\langle f^+_if_i\rangle $),
the local density of light atoms ($n^d_i=\langle d^+_id_i\rangle $),
the total site occupation ($n_i= \langle f^+_if_i+d^+_id_i \rangle $),
the variance of the local density ($\Delta_i=\langle (f^+_if_i+d^+_id_i)^2\rangle - 
\langle f^+_if_i+d^+_id_i\rangle ^2$) and the local double occupation 
($D_i=\langle d^+_id_if^+_if_i\rangle$), as functions of total number of the 
confined atoms, the Coulomb interaction $U$ 
and the confining potential $V$. The ground states are calculated by 
a well-controlled numerical method that we have elaborated~\cite{Fark1} for 
a description of the conventional Falicov-Kimall model ($V=0$). 
Later, the method was successfully used for various generalizations 
of the Falicov-Kimball model and different physical
problems~\cite{Fark2}. Its 
generalization on systems with a harmonic potential is straightforward too.
\section{Results and discussion}

In Fig.~1 we present results of our numerical calculations for $n_i^f, n^d_i,
n_i, D_i$ and $\Delta_i$ obtained on the one-dimensional cluster of $L=120$ sites 
at $U=4$, $V=4$ and different fillings. We have added also the profiles of 
the local compresibility that has been proposed by Rigol et al.~\cite{Rigol1} 
as a local order parameter to characterize the Mott-insulator regions. 
This quantity is defined as~\cite{Rigol1}
\begin{equation} 
\kappa_i^{\it l}=\sum_{|j|\leq{\it l}(U)} \chi_{i,i+j}\ ,
\label{eq2}
\end{equation} 
where
\begin{equation}
\chi_{i,j}=\langle n_in_j\rangle-\langle n_i\rangle\langle n_j\rangle
\label{eq3}
\end{equation}
is the density-density correlation function and ${\it l}(U)\sim b\xi(U)$, with 
$\xi(U)$ the correlation length of $\chi_{i,j}$ in the unconfined system at 
half-filling for a given $U$. The factor $b$ is chosen within a range where 
$\kappa^{\it l}$ becomes qualitatively insensitive to its precise
value~\cite{Rigol2}. The insulating regions are then characterized
by  $\kappa_i^{\it l}=0$. For the values of $U$ used here we usually have
$b\sim 4$-$8$ with $\xi(U) \sim 1$. 

The most interesting result obtained at low particle fillings (we note that
$N_f=N_d=N/2$) is the observation of the complete phase separation with heavy 
atoms in the center of the trap ($n^f_i=1, n_i^d \sim 0$) and light atoms 
in the surrounding regions ($n^f_i=0, n_i^d<1$). Such a behaviour is found 
for all particle fillings from $N_f=1$  to some critical value $N^c_f$ that 
rapidly decreases with increasing $V$ and is almost independent of the local 
Coulomb interaction $U$. In the regions where $n^f_i=0$ ($0 < n^d_i < 1$) the 
variance of the local density $\Delta_i$ and the local compresibility 
$\kappa_i^{\it l}$ are finite (the metallic phase), 
while in the middle of the trap where $n_i^f=1$ ($n_i^d$ exponentially
decreases in this region) both $\Delta_i$ and $\kappa_i^{\it l}$
are equal to zero (the insulating phase). Thus in accordance with results 
obtained for the Hubbard model~\cite{Rigol1} (the hopping probabilities are 
same for both types of atoms) we have found that also in the Falicov-Kimball 
model insulating domains coexist with metallic regions, such that global 
quantities are not appropriate to describe the system. 

At higher particle fillings the situation is more complex. Above the critical 
filling $N^c_f$ ($N^c_f=24$, for $U=4,V=4$) the connected cluster by heavy atoms occupied sites (in the center 
of the trap) splits on smaller clusters, usually of the same size, separated by
the empty site. As $N_f$ increases the size of clusters decreases from $N_f$ to 1. 
Of course, the redistribution of heavy atoms has dramatic consequences on the 
distribution of light atoms. Now, the light atoms occupy preferably the empty 
sites in the middle
of the trap ($n_i \sim 1$) what leads to the atomic-density waves in $n_i^d$ 
and $n_i$ profiles. In the region where $n_i\sim 1$ the variance of the
local density is finite but smaller 
than in surrounding metallic regions indicating~\cite{Rigol1,Rigol2} 
the presence of Mott-insulating 
phase in the center of the trap. This conjecture supports the behaviour of the 
local compresibility $\kappa^{\it l}_i$ that is equal to zero in the 
corresponding region.

Increasing the number of particles up to $N_f=52$, the connected cluster of heavy 
atoms ($n_i^f=1$) starts to form in the middle of the trap. In this region both 
the variance of the local density and the local compresibility are finite what 
indicates the presence of metallic phase in the center of the trap. 
Upon adding more fermions, this 
new metallic phase widens spatially, while the Mott-insulating regions of the atomic 
density waves are pushed to the borders and completely disappear at $N_f=90$. At this 
filling practically the whole region of the trap, except the edges (where 
$n_i^f=0, n_i^d\sim 1$), is metallic. With still a higher filling this metallic phase 
is further stabilized, but at some critical filling ($N_f=95$) a new insulating 
phase ("a band insulator") starts to develop in the center of the trap 
($n_i^f=1, n_i^d=1$). This trend holds also for the highest particle fillings,
the width of the band-insulating phase increases and the surrounding metallic 
regions are gradually suppressed.

To reveal the role of the Coulomb interaction $U$ and the confining
potential $V$ on a formation of metallic and insulating domains, similar
calculations have been also performed for various combinations of $U$ and 
$V$. The results of numerical calculations for the site occupation $n_i$
as a function of $i$ are displayed in Fig.~2. One can see that the Coulomb
interaction $U$ and the confining potential $V$ exhibit precisely opposite
effects on the stability of metallic and insulating domains. Indeed, with
increasing $V$ (at fixed $U$) the insulating domain $(n_i\sim 1)$ is 
suppressed and the metallic domains are stabilized, while with 
increasing $U$ (at fixed $V$) the metallic domains are suppressed 
and the insulating domain is stabilized.    
 
Since the Falicov-Kimball model can be considered as a simplification of 
the Hubbard model (only one kind of particles, say with spin up can hop) it is 
interesting to compare results obtained in these two different limits. Such a 
comparison (see Ref.~3 and Ref.~4) reveals obvious differences in behavior 
of these models in the confining potential. For example, the ground-state of 
the Hubbard model (for $V\neq 0$) is always metallic at low particle fillings,
while the metallic regions coexist with the insulating region in the ground 
state of the Falicov-Kimball model. Moreover, the local density profiles 
exhibit obvious oscillations for the Falicov-Kimball model, while 
no sign of such oscillations has been observed for the repulsive Hubbard model 
(with the exception of Hartree-Fock~\cite{Rigol2} and variational 
studies~\cite{Fujihara}, that were not confirmed by a projector Monte Carlo 
simulations, however~\cite{Rigol2}).

To exclude the possibility that oscillations are a consequence of a finite size
of clusters used in our numerical calculations, we have performed an exhaustive
finite-size scaling analysis on finite clusters up to $L=480$ sites for all 
particle fillings from Fig.~1. This analysis showed that the ground states
found for $L=120$ hold also on clusters of $L=240$ and $L=480$ sites and thus 
they can be satisfactorily extrapolated on much larger clusters. In Fig.~3
we present numerical results for extrapolated ground states obtained for
a cluster of $L=6000$ sites and the same values of $n_f=N_f/L, V$ and $U$ as used 
in Fig.~1. These results clearly demonstrate that increasing $L$ suppresses
the atomic-density oscillations in the metallic phase, but stabilizes
the atomic-density waves in the insulating phase.
   
Since in the real experiments with ultracold atoms the hopping matrix elements 
(of heavy atoms) between the nearest-neighbor sites ($t_f$) are not strictly 
equal to zero, it is necessary to examine the stability of our solutions obtained 
for $t_f=0$ against the finite values of $t_f$. For this reason we have performed 
exhaustive exact-diagonalization studies of the asymmetric Hubbard 
model~\cite{asy} ($t_f\neq 0$) in a confining potential for a wide range of 
model parameters ($N_f, t_f, U$ and $V$) on finite clusters up to 
$L=12$. The representative examples of local density profiles are displayed in 
Fig.~4 for several different values of $t_f$. They clearly show that results 
obtained for $t_f=0$ remain stable also at finite (small) $t_f$. Thus the 
Falicov-Kimball model with a confining potential can be used satisfactorily 
to model the ground-state behavior of mixtures of two-component fermionic 
atoms (with strongly different masses) in a harmonic potential.

Finally, we have also calculated the variance of the local density as a 
function of the local density $n_i$. It is known from the study of the 
one-dimensional Hubbard model in a harmonic potential~\cite{Rigol1} 
that this quantity shows universality with respect to the confining potential 
for $n_i\to 1$. It should be noted that the universal behavior was observed only 
for strong Coulomb interactions $U$, where systems have a Mott-insulating 
phase at $n=1$. In Fig.~5 we present numerical results for the variance of the
local density obtained within the Falicov-Kimball model with a harmonic potential as 
a function of the local density calculated for various particle fillings and various 
values of $U$ and $V$. One can see that all numerical data for the variance of
the local density collapse on the same curve (given by $\Delta=-n(n-1)$ 
for $n \leq 1$ and $\Delta=-(n-1)(n-2)$ for $n > 1$). Thus in contrast 
to the similar studies on the Hubbard model~\cite{Rigol1} we have found that the 
variance of the local density of systems 
described by Falicov-Kimball model with a harmonic potential exhibits universality 
not only with respect to the confining potential but also with respect to the 
Coulomb interaction $U$. In addition, we have found that the universality hold not
only for $n_i\to 1$, but for all values of $n_i$. A similar universal 
behaviour has been observed also for other local quantities. For example, 
the local double occupation $D_i=0$ for $n_i \leq 1$ and $D_i=n_i-1$ for 
$n_i > 1$, independent of values of $U,V,N_f$ and $L$.  
Also the universal behavior is found for the local compresibility
$\kappa^{\it l}$ when $n_i\to 1$. However, the value of the critical exponent
for the Falicov-Kimball model is equal to 1, unlike the nontrivial value 
0.68-0.78 found for the Hubbard model~\cite{Rigol1}.

In summary, we have studied the ground-state properties of fermionic mixtures 
with mass imbalance in a one-dimensional optical lattice within 
the spinless Falicov-Kimball model with a harmonic potential. 
We have found that the system exhibits the phase separation at low particle 
fillings. In this case the heavy atoms occupy the center of the trap 
while the light atoms are localized in the surrounding (metallic) regions. 
At higher fillings we have observed a formation of Mott-insulating domains
with atomic-density waves. In all cases Mott-insulating phases coexist 
with metallic phases. One of the most interesting results is, however,
the observation of the universal behavior of the variance of the local 
density (independent of the particle filling, the Coulomb interaction and the
strength of the confining potential) over the whole region of the local density 
values.

\vspace*{1cm}
This work was supported by Slovak Grant Agency VEGA under Grant
No.2/7057/27 and Slovak Research and Development Agency (APVV) under
Grant LPP-0047-06.

\newpage
\centerline{\bf Figure Caption}

\vspace{0.5cm}
Fig.~1. Profiles of various local quantities along the trap calculated for 
$V=4, U=4, L=120$ and four different values of particle fillings. 
Different panels (from top to bottom) correspond to: the local density 
of heavy atoms $n^f_i$, the local density of light atoms $n^d_i$, the total 
site occupation $n_i=n^f_i+n^d_i$, the local double occupation $D_i$, 
the variance of the local density $\Delta_i$ and the local compresibility 
$\kappa^{\it l}_i$.

\vspace{0.5cm}
Fig.~2. The site occupation $n_i$ as a function of site position $i$ 
calculated for different $V$ (the first column) and different $U$
(the second column).

\vspace{0.5cm}
Fig.~3. The site occupation $n_i$ as a function of site position $i$ 
calculated for extrapolated ground states for different particle fillings. 
The parameters involved are $V=4, U=4$ and $L=6000$. The insets show 
details of $n_i$ profiles for the region of atomic-density waves.

\vspace{0.5cm}
Fig.~4. The exact-diagonalization results (the asymmetric Hubbard model 
with a harmonic trap) for the site occupation $n_i$ as 
a function of site position $i$ calculated for different values of 
$t_f, N_f, U$ and $V$ on a finite cluster of $L=12$ sites.

\vspace{0.5cm}
Fig~5. The variance of the local density $\Delta$ as a function of the 
local density $n$ calculated for different values of $U, V$ and $N_f$. 
The numerical data are fitted by $\Delta=-n(n-1)$ for $n \leq 1$
and $\Delta=-(n-1)(n-2)$ for $n > 1$. 

\newpage
\begin{thebibliography}{99}

\bibitem{Greiner} M. Greiner et al., Nature (London) {\bf 415}, 39 (2002); 
I. Bloch, Phys. World {\bf 17}, 25 (2004); D. Jacksch and P. Zoller, 
Ann.Phys. (N.Y.) {\bf 315}, 52 (2005); W. Hofstetter, Philos. Mag. {\bf 86},
1891 (2006); T. Stoferle et al., Phys. Rev. Lett. {\bf 96}, 030401 (2006).

\bibitem{Cirac} J.I. Cirac and P. Zoller, Science {\bf 301}, 176 (2003);

\bibitem{Rigol1} M. Rigol, A Muramatsu, G.G. Batrouni and R.T. Scalettar, 
Phys. Rev. Lett. {\bf 91}, 130403 (2003).

\bibitem{Rigol2} M. Rigol and A Muramatsu, Phys. Rev. A. {\bf 69}, 053612
(2004).

\bibitem{Ates} C. Ates and K. Ziegler, Phys. Rev. A. {\bf 71}, 063610
(2004); K. Ziegler, Nuclear Physics A {\bf 790}, 718C (2007).

\bibitem{Gu} S. J. Gu, R. Fan, and H. Q. Lin,  Phys. Rev. B {\bf 76},
125107 (2008).
                                                                                
\bibitem{Caza} M.A. Cazalilla, A.F. Ho and T. Giamarchi, 
Phys. Rev. Lett. {\bf 95}, 226402 (2003).

\bibitem{Falicov} L.M. Falicov and J.C. Kimball, Phys. Rev. Lett.
{\bf 22}, 997 (1969).

\bibitem{Gottwald} T. Gottwald and P.G.J van Dongen, Eur. J. Phys.
{\bf 61}, 277 (2008).

\bibitem{Gruber} 
P. Lemberger, J. Phys. A {\bf 25}, 715 (1992);
C. Gruber and D. Ueltschi and J. Jedrzejewski, 
J. Stat. Phys. {\bf 76}, 125 (1994); J.K. Freericks, Ch. Gruber and 
N. Macris, Phys. Rev. B {\bf 53}, 16189 (1996); J.K. Freericks, 
E.H. Lieb and D. Ueltschi, Phys. Rev. Lett. {\bf 88}, 106401 (2002).

\bibitem{Fark1} P. Farka\v{s}ovsk\'y, Eur. J. Phys. B {\bf 20}, 209 (2001);
P. Farka\v{s}ovsk\'y, Int. J. Mod. Phys. B {\bf 17}, 4897 (2003).

\bibitem{Fark2} P. Farka\v{s}ovsk\'y, H. \v{C}en\v{c}arikova and
N. Toma\v{s}ovi\v{c}ova, Eur. J. Phys. B {\bf 45}, 479 (2005);
P. Farka\v{s}ovsk\'y and  H. \v{C}en\v{c}arikova, Eur. J. Phys. B {\bf 47},
517 (2005).

\bibitem{Fujihara} Y. Fujihara, A. Koga and N. Kawakami, 
J. Phys. Soc. Jpn.B {\bf 76}, 034716 (2007).

\bibitem{asy} R. Lyzwa and  Z. Domanski, Phys. Rev. B {\bf 50}, 11381 (1994); 
P. Farka\v{s}ovsk\'y, Phys. Rev. B {\bf 77}, 085110 (2008).


\end{thebibliography}
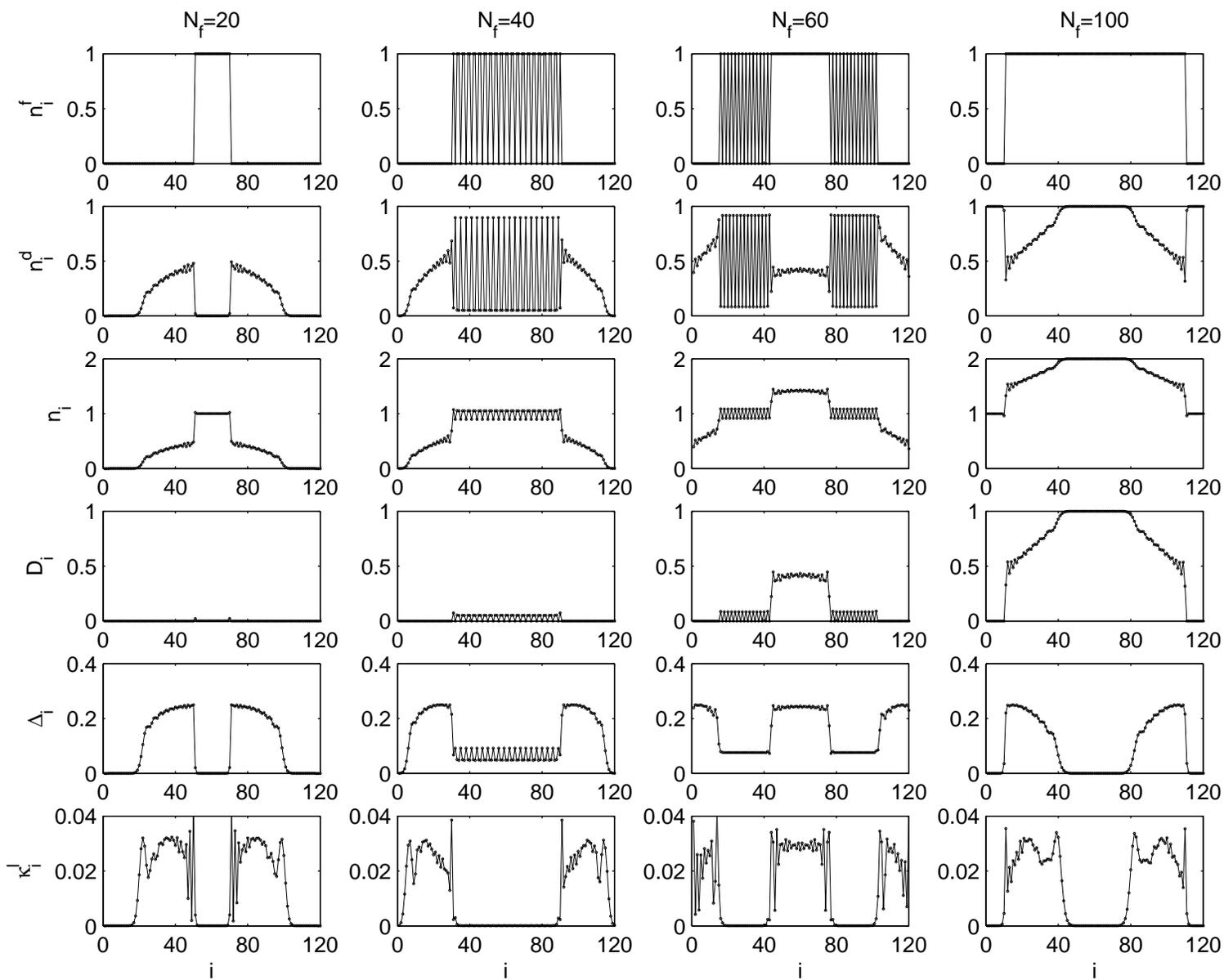
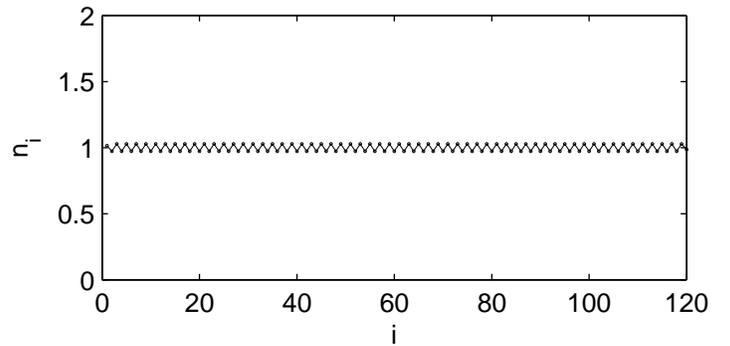
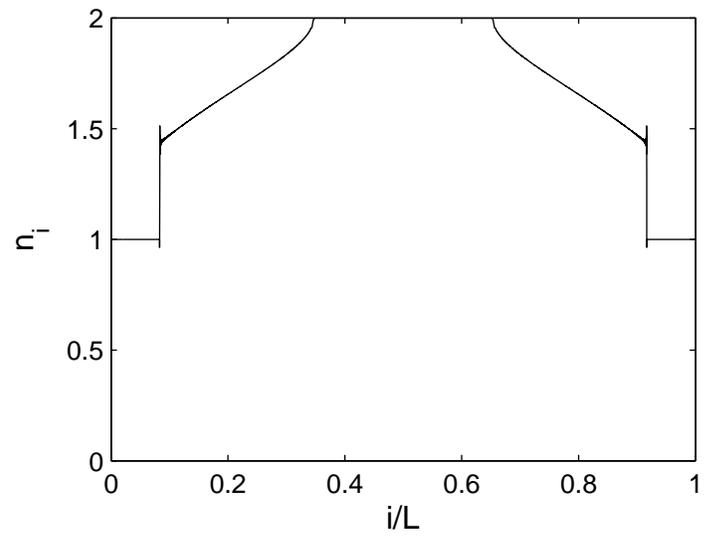
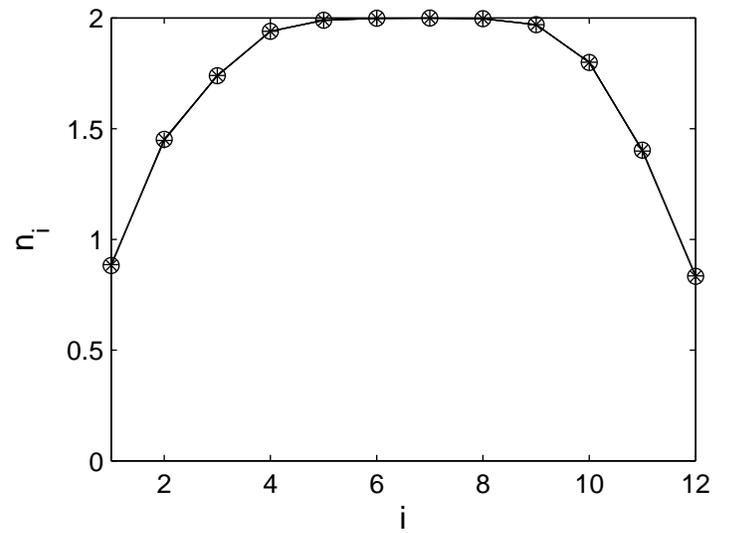
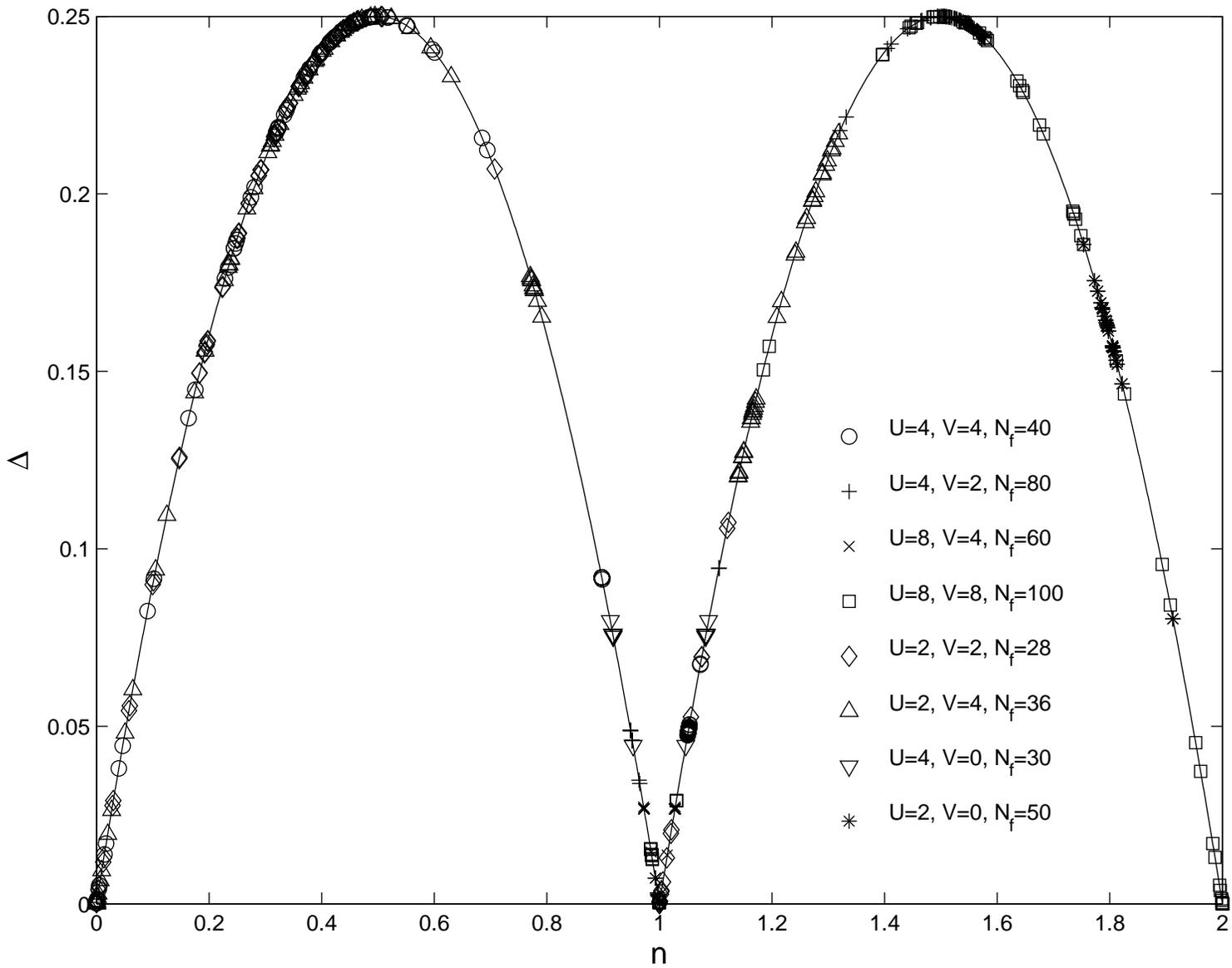
\end{document}